\documentclass[%
 reprint,
 amsmath,amssymb,
 aps,
]{revtex4-1}

\usepackage{graphicx}
\usepackage{dcolumn}
\usepackage{bm}

\begin{document}


\title{Force-free Currents and the Newman-Penrose Tetrad of a Kerr Black Hole: Exact Local Solutions}

\author{Govind Menon}
\affiliation{Department of Chemistry and Physics\\ Troy University, Troy, Al 36082}

\date{\today}

\begin{abstract}
In a previous article we derived a class of solutions to the force-free magnetosphere in a Kerr background. Here, the streaming surface, defined by constant values of the toriodal component of the electromagnetic vector potential $A$, were generated by constant values of $\theta$. The electromagnetic current vector flowed along the in-falling principle null geodesic vector of the  geometry. Subsequently, we generalized this to obtain an out-going principle null geodesic vector as well. In this article, we derive solutions that are complimentary to the above mentioned criteria. Namely, here the solution has a streaming surface generated by spheres of constant radial coordinate $r$, and the current vector is generated by linear combinations of $m$ and $m^\star$, the remaining bases vectors in the Newman-Penrose null tetrad.
\end{abstract}

\pacs{Valid PACS appear here}
\maketitle


\section{INTRODUCTION}

In 1977, Blandford and Znajek argued for  force-free electrodynamics  as a reasonable mathematical framework describing a black hole magnetosphere   (\cite{BZ77}). The force-free magnetosphere in a Kerr background has received considerable attention off late due to its expected properties with regards to its ability to extract energy and angular momentum from a rotating black hole in an astrophysical setting (for example see \cite{TT14}, \cite{GT14}, \cite{ZYL14} and \cite{LRS14}).
The basic mechanism we expect is that of a plasma interacting with its own electromagnetic field near the vicinity of the event horizon of the black hole. The plasma relaxes, by possibly ejecting helicity, thereby producing a force-free magnetosphere. Jet formation in radio-loud, active galactic nuclei could be explained if indeed force-free electrodynamics permitted solutions with the desired features.

In \cite{MD11} we were successful in creating an exact solution that allowed the extraction of energy via a electromagnetic poynting flux. Although, mathematically consistent, it was not clear how this solution could be physically realized in nature. The possible decomposition of the net electromagnetic current into leptonic flows and a baryonic fluid remains an unanswered  question. If the actual ejection of current and energy happens at the event horizon, it is necessary to match an outgoing exterior solution with a relevant infalling interior solution. In this case, the interior fields must satisfy the Znajek regularity condition at the event horizon \cite{ZNA77}.

In this paper however, we have a different purpose. In \cite{MD07} we obtained the first exact solution to the force-free magnetosphere in a Kerr background. The resulting current vector was proportional to the in-falling principle null geodesic vector field
\begin{equation}
n= \frac{r^2+a^2}{\Delta}\;\partial_t- \partial_r +\frac{a}{\Delta}\;\partial_\varphi.
\label{ndef}
\end{equation}
The vector field above is written in the standard Boyer-Lindquist coordinates. This corresponding solution did satisfy the Znajek regularity condition and hence was well defined at the event horizon of the black hole. Using the symmetries of the equations of electrodynamics in the case of axis-symmetric and stationary solutions, we were successful in finding a similar solution wherein the current vector was proportional to the out-going null vector field of the Newman-Penrose tetrad
\begin{equation}
l= \frac{r^2+a^2}{\Delta}\;\partial_t+ \partial_r +\frac{a}{\Delta}\;\partial_\varphi.
\label{ldef}
\end{equation}
However, unlike the previous case, this new solution was not valid at the event horizon \cite{MD11}.
In the sections below, we will now construct two local, linearly independent, exact solutions to the force-free magnetosphere, wherein the current vectors are proportional to the linear combinations of $m$ and $m^\star$ (the complex conjugate of $m$) where
$$m = \frac{1}{\sqrt{2} \rho} \left(i a \sin \theta, \;0, \;1,\; \frac{i}{\sin \theta}\right).$$
These new  solutions are local in the sense that they are not strictly valid at the event horizon and the symmetry axis of the Kerr black hole. The point to also note is that the Newman-Penrose tetrad in a Kerr background are precisely the vectors $n, l, m$ and $m^\star$. We will have then succeeded in establishing a direct connection between fore-free electrodynamics and the Newman-Penrose tetrad.

After presenting a brief overview of force-free electrodynamics in a Kerr background in the next section, we move onto presenting our new solution in section \ref{omchoices}. We then conclude by discussing the immediate properties of this new solution.

\section{Basic Equations of Force-free Dynamics}
In the Boyer-Lindquist coordinates, the Kerr metric takes the form:
$$ds^2=( \beta^2 - \alpha^2 )\;dt^2 \;+ $$
\begin{equation}
 \;2 \;\beta_\varphi \;d\varphi
\;dt
+\gamma_{rr}\; dr^2 + \;\gamma_{\theta \theta}\; d\theta^2 +
\;\gamma_{\varphi\varphi}\;d\varphi^2 \;,
\end{equation}
where the metric coefficients are given by
$$\beta_\varphi \; \equiv g_{t \varphi}\; = \;\frac{-2Mr a
\sin^2\theta}{\rho^2}\;,\;\;\;\gamma_{rr} =
\frac{\rho^2}{\Delta}\;,$$
$$\beta^2-\alpha^2 \;= \;g_{tt} \;=\; -1 + \frac{2Mr}{\rho^2}\;,$$

$$
\gamma_{\theta \theta} = \rho^2, \;\; {\rm and}\;\; \gamma_{\varphi \varphi} = \frac{\Sigma^2 \sin^2\theta}{\rho^2}\;.
$$
Here,
$$\rho^2 = r^2 + a^2
\cos^2\theta\;,\;\;\;\Delta = r^2 -2 M r + a^2$$
and
$$
\Sigma^2 = (r^2 + a^2)^2 -\Delta \; a^2 \sin^2\theta\;.
$$
Also
$$
\alpha^2 = \frac{\rho^2 \Delta}{\Sigma^2}, \;\;\; \beta^2 =
\frac{\beta_\varphi^2}{\gamma_{\varphi \varphi}}\;,\;\sqrt{\gamma} = \sqrt{\frac{\rho^2\;\Sigma^2}{\Delta}}\;\sin \theta\;,$$
and
$$\sqrt{-g}=\alpha\; \sqrt{\gamma} = \rho^2 \sin\theta\;.$$

Maxwell's equations in any spacetime can be written as
\begin{equation}
\nabla_\beta \star \;F^{\alpha \beta} = 0 \;, \;{\rm and} \;
\nabla_\beta  F^{\alpha \beta} = I^\alpha\;.
\label{maxeq}
\end{equation}
Here $F^{\alpha \beta}$ is the Maxwell stress tensor,  $I^\alpha$ is the  electric current  density vector and $\nabla$ is the covariant derivative of the geometry. $\star \;F$ is
the two form defined by
\begin{equation}
\star \;F^{\alpha \beta} \equiv \frac{1}{2}\epsilon^{\alpha \beta \mu \nu} F_{\mu \nu}\;.
\end{equation}
Here, $\epsilon_{\alpha \beta \mu \nu}$ is the completely antisymmetric Levi-Civita tensor density of spacetime such that $
\epsilon_{0123}= \sqrt{-g}=  \alpha \sqrt{\gamma}
$. In this paper, we choose to work with electric and magnetic fields rather than the covariant formalism of electrodynamics. Thus we define
$ E$ and $ B$ vector fields by
\begin{equation}
 F_{ \mu  \nu} =
\left[\begin{array}{cccc}
0&  -E_1&  -E_2 & -E_3\\
  E_1 & 0 & \sqrt{\gamma}\; B^3 & - \sqrt{\gamma}\; B^2\\
E_2& - \sqrt{\gamma} \;B^3 & 0 & \sqrt{\gamma} \;B^1\\
E_3 & \sqrt{\gamma}\;B^2 & - \sqrt{\gamma} \;B^1 & 0\\
\end{array}\right] \;.
\label{fdown}
\end{equation}
In the remainder of this section, we simply summarize the equations of force-free dynamics in a 3+1 formalism. The details behind the calculations can be found in \cite{MD05} and \cite{K04}. When, as in the case of the Kerr metric in Boyer-Lindquist coordinates, $\partial_t$ corresponds to the Killing vector that generates time translations, eq.(\ref{maxeq}) become the familiar set of Maxwell equations with the following modifications:
\begin{equation}
\tilde \nabla \cdot B = 0\;,
\label{divb}
\end{equation}
\begin{equation}
\partial_t B + \tilde \nabla \times E = 0\;,
\label{faraday}
\end{equation}
\begin{equation}
\tilde \nabla \cdot D = \rho\;,
\label{maxcharge}
\end{equation}
and
\begin{equation}
-\partial_t D + \tilde \nabla \times H = J\;.
\label{maxcurrent}
\end{equation}
Here, $\rho = \alpha I^t$ and $J^k = \alpha I^k$, also,
\begin{equation}
\alpha D = E- \beta \times B
\label{contitutive1}
\end{equation}
and
\begin{equation}
H = \alpha B - \beta \times D\;.
\label{contitutive2}
\end{equation}
The generalized cross product of vector fields are defined by
\begin{equation}
(A \times B)^i \equiv \; \epsilon^{ijk}  \; A_j\; B_k\;,
\end{equation}
and
$\tilde \nabla$ corresponds to the induced covariant derivative of the 3-D absolute space defined by surfaces of constant $t$. Now we impose the requirements of force-free electrodynamics. By definition, here
\begin{equation}
F_{\nu \alpha}\; I^\alpha =0\; .
\label{divtemforce}
\end{equation}
In the 3+1 formalism, this condition reduces to
\begin{equation}
E \cdot J = 0
\label{fofree1}
\end{equation}
and
\begin{equation}
\rho E + J \times B = 0.
\label{fofree2}
\end{equation}
Eqs. (\ref{divb}) through (\ref{fofree2}) define the entire content of what we mean by force-free electrodynamics in a Kerr background.
Finally, for the electromagnetic field to be well defined at the event horizon of the Kerr black hole, the following Znajek regularity condition should be satisfied (\cite{ZNA77}):
\begin{equation}
H_\varphi \left|_{r_+} = \frac{\sin^2\theta}{\alpha}\; B^r\; (2Mr\; \Omega -a) \right |_{r_+}.
\label{znaregcond}
\end{equation}
Now we impose the symmetries of the background metric onto the electromagnetic fields and currents. Specifically, we require that all electrodynamics quantities are time-independent and axis-symmetric. It is then not difficult to show that there exists a vector $\omega =\Omega \partial_\varphi$ such that
\begin{equation}E = - \omega \times B\;.
\label{omdefeqn}
\end{equation}
The vanishing of the curl of $E$ for time-independent solutions gives that
$$\tilde \nabla_B \Omega =0\;.$$
Therefore, the magnetic field can be written as
\begin{equation}
B_P = \frac{\Lambda}{\sqrt\gamma}\left(-\Omega_{,\theta}  \;\partial_r + \Omega_{,r}\;
 \partial_\theta \right)\;.
\label{bpexplicit}
\end{equation}
The subscript $P$ on $B$ indicates the poloidal ($r$ and $\theta$) components of $B$. In standard notation
$$\Lambda = -\frac{d A_\varphi}{d\Omega}\;,$$
where $A_\varphi \equiv A_T$ is the toroidal ($\varphi$ component) of the vector potential $A$. Clearly, this is possible only because $A_\varphi$ depends only on the value of $\Omega$. Eq.(\ref{omdefeqn}) implies that $E_T=0$, and $E_P \cdot B_P= 0$. On the other hand, for force-free dynamics $ E_P \cdot J_P =0$. Thus, we can conclude that $ B_P \propto J_P$. This along with eq.(\ref{maxcurrent}) implies that the toroidal component of $H$ must satisfy
\begin{equation}
\tilde \nabla_B H_\varphi =0\;.
\label{hphistream}
\end{equation}
I.e.,
$H_\varphi$ is also dependent only on the value of $\Omega$.

Eq.(\ref{fofree2}) is trivial when projected to plane spanned by $\partial_\varphi$ and $B_P$. However, when projected along $E_P$, eq.(\ref{fofree2}) gives the only remaining master constraint equation for time-independent, axis-symmetric, force-free electrodynamics.
The resulting constraint equation reduce to a manageable form when written in terms of the streaming function $\Omega$. To facilitate this define
\begin{equation}
\chi(\Omega)=\partial_t +\omega =\partial_t + \Omega\; \partial_\varphi.
\label{kerrcanvf}
\end{equation}
Let
\begin{equation}
\chi_\varphi (\Omega) \equiv g (\partial_t + \Omega\; \partial_\varphi,\partial_\varphi) = g_{t\varphi} + \Omega \; g_{\varphi \varphi}
\label{kerrgeospc1}
\end{equation}
and
\begin{equation}
\chi_t (\Omega)\equiv g(\partial_t + \Omega\; \partial_\varphi,\partial_t)= g_{t t} + \Omega \; g_{t \varphi}\;.
\label{kerrgeospc2}
\end{equation}
In terms of $\Omega$ and $\Lambda$
$$(\rho E + J \times B ) \cdot E_p= 0$$
is satisfied provided
\begin{widetext}
\begin{equation}
\frac{1}{2 \Lambda } \frac{d H_\varphi^2}{d \Omega}=\alpha \gamma_{\varphi \varphi} \Omega \;\tilde \nabla \cdot \left(\frac{\Lambda}{\alpha \gamma_{\varphi \varphi}} \chi_\varphi (\Omega) \tilde \nabla\Omega\right)+
\alpha \gamma_{\varphi \varphi} \tilde \nabla \cdot \left(\frac{\Lambda}{\alpha \gamma_{\varphi \varphi}} \chi_t (\Omega) \tilde \nabla\Omega\right)\;,
\label{consteqlong1}
\end{equation}
or equivalently
\begin{equation}
\frac{1}{2 \Lambda } \frac{d H_\varphi^2}{d \Omega}=\alpha \gamma_{\varphi \varphi} \tilde \nabla \cdot \left(\frac{\Lambda}{\alpha \gamma_{\varphi \varphi}} \chi^2(\Omega) \tilde \nabla\Omega\right)-\Lambda \chi_\varphi (\Omega) (\tilde \nabla \Omega)^2\;.
\label{consteqlong2}
\end{equation}
Here $\chi^2(\Omega)= g(\chi(\Omega), \chi(\Omega))$ and $(\tilde \nabla \Omega)^2= \gamma^{ij}\Omega_{,i}\Omega_{,j}$.
\end{widetext}

The main point and difficulty here is that, just like the left hand side of the equation above (see eq.(\ref{hphistream})), the right hand side of the equation above must depend only on the streaming function $\Omega$. The functions $\Lambda$ and $\Omega$ determine every other electromagnetic quantity uniquely.

\section{Canonical choices of $\Omega$}
\label{omchoices}
Due to stationarity and axis-symmetry, the Kerr metric mixes the standard Boyer-Lindquist  $t$ and $\varphi$ coordinates in a particular simple way:

$$g_{tt} + \frac{1}{a  \sin^2 \theta } \;g_{t \varphi} = -1 ,$$
$$g_{t \varphi} + \frac{1}{a  \sin^2 \theta } \;g_{\varphi \varphi} = \frac{r^2+a^2}{a} ,$$
and
$$ g_{t \varphi} + \frac{a}{r^2 + a^2}\; g_{\varphi \varphi} =  \frac{a}{r^2+a^2}\; \sin^2 \theta \Delta ,$$
$$g_{t t} + \frac{a}{r^2 + a^2}\; g_{t \varphi} =  - \frac{a}{r^2 + a^2} \; \Delta .$$
Thus, there are two special choices for $\Omega$ in Kerr geometry in the Boyer-Lindquist coordinate system that makes $\chi_t$ and $\chi_\varphi$ exceptionally simple (i.e., written in terms of well known functions of the spacetime geometry). However, {\it a priori}, there is no guaranty that these choices are consistent with eq.(\ref{consteqlong1}).

\subsection{Recovering Our Previous Solution}
When
$$
\Omega \equiv \Omega_1 = \frac{1}{a \sin^2 \theta},
$$
we have that
$$\chi_t (\Omega_1) = -1$$
and
$$\chi_\varphi (\Omega_1) = \frac{r^2 + a^2}{a}\;.$$
In \cite{MD07}, we have shown that this is indeed a valid choice for $\Omega$ since in this case
$$\frac{1}{2 \Lambda } \frac{d H_\varphi^2}{d \Omega}=\frac{-2}{a^3 \sin\theta} \;\frac{d}{d\theta}\left[\frac{\Lambda \cos\theta}{\sin^4 \theta}\right],$$
and the right hand side above,
like $\Omega$, is only a function of $\theta$ for any arbitrary but sufficiently regular $\Lambda(\theta)$ (so as to make the solutions well defined at the poles). This can be easily integrated to give
\begin{equation}
H_\varphi=\alpha B_{\varphi} =\frac{2}{a^2} \Lambda \frac{\cos\theta}{\sin^4\theta}.
\label{hphi_1}
\end{equation}
It also easily checked that the above equation satisfies the Znajek regularity condition. The other non-trivial components of the electric and magnetic fields are
\begin{equation}
B^r=\frac{2}{a} \Lambda \frac{\cos{\theta}}{\sqrt{\gamma}\sin^3{\theta}}.
\label{radb_1}
\end{equation}
\begin{equation}
E_\theta =-\frac{2}{a^2} \Lambda \frac{\cos{\theta}}{\sin^5{\theta}}.
\end{equation}
The electromagnetic current in this configuration is given by
\begin{equation}
I =-\frac{2}{a^2\; \alpha\; \sqrt\gamma} \;\;\frac{d}{d \theta}\left[\Lambda \frac{\cos\theta}{\sin^4\theta}\right]n,
\end{equation}
where $n$ is the infalling principle null geodesic of the Kerr geometry explicitly given in eq.(\ref{ndef}).
In \cite{MD11}, we showed that the inherent assumed symmetries of the fields (axis-symmetry and time- independence) implied that there exists a dual solution with the same $\Omega = \Omega_1$ such that the current vector was proportional to the outgoing principle null geodesic of the kerr geometry $l$ (see eq.(\ref{ldef})).
Here the explicit forms of the electromagnetic field components are not very different except for a few sign changes (see \cite{MD11} for details). This dual solution however does not satisfy the Znajek regularity condition. Here, the surfaces of constant $\Omega$ are cones (surfaces of constant Boyer-Lindquist coordinate $\theta$). Recently, the solution presented here was extended to the  $\varphi$ dependent case by Brennan et. al. (see \cite{BGJ13}).

\subsection{A New Class Of Solutions}
Now consider a second canonical choice
$$
\Omega \equiv \Omega_2 (r) = \frac{a}{r^2 +a^2}.
$$
In this case, we have that
$$\chi_\varphi (\Omega_2) = \frac{a}{r^2 + a^2} \sin^2\theta \Delta$$
and
$$\chi_t (\Omega_2)=-\frac{\Delta}{r^2 + a^2}\;.$$
In this case, once again we find that the right hand side of eq.(\ref{consteqlong1}) becomes
$$
\frac{1}{2 \Lambda } \frac{d H_\varphi^2}{d \Omega} = -\frac{\Delta \Omega }{a^2} \frac{d}{dr}\left[\Lambda \Delta \Omega \Omega_{,r}\right].$$
Clearly, the right hand side above,
like $\Omega$, is only a function of $r$ for any arbitrary but sufficiently regular $\Lambda(r)$. This is also easily integrated to give
\begin{equation}
a H_\varphi = a \alpha B_\varphi =\pm \sqrt{C^2 - (\Lambda \Delta \Omega \Omega_{,r})^2}\;.
\label{rdephphi}
\end{equation}
Here $C$ is an integration constant large enough to ensure that $H_\varphi$ is real.
Eq.(\ref{bpexplicit}) gives the following explicit solution for the poloidal components of the magnetic field:
$$B_P = \frac{-2 a r \Lambda}{\sqrt{\gamma}\; (r^2+a^2)^2} \;\partial_\theta.$$
Here the only non-zero component of the electric field given by eq.(\ref{omdefeqn}) takes the form
$$E_P= \frac{-2 a^2 r \Lambda}{(r^2+a^2)^3}\; dr.$$
The components of the dual fields ($D$ and $H$) are given by
$$D= D_P = \frac{\Lambda}{\alpha \gamma_{\varphi \varphi}} \;\chi_\varphi (\Omega) \;d\Omega$$
which in our case reduces to
$$D= -\frac{\Lambda}{\alpha \gamma_{\varphi \varphi}} \;\frac{2 r a^2 }{(r^2+a^2)^3} \;\sin^2\theta \Delta\; dr\;.$$
Also, in general
$$H_P = -\chi_t (\Omega) \;\frac{B_P}{\alpha}$$
which for our particular choice of $\Omega_2$ reduces to
$$H_P = -\frac{2 a r \Lambda}{ (r^2+a^2)^3}\;\frac{\Delta}{\rho^2 \sin \theta} \;\partial_\theta.$$
It is important to note that there is still a choice in the explicit form for $\Lambda$. Here, the surfaces of constant $\Omega$ are spheres (surfaces of constant Boyer-Lindquist coordinate $r$).
\section{Discussion And Conclusion}
For the remainder of the section, we restrict our discussion the solution generated by
$\Omega= \Omega_2 (r)$.
Since $B^r =0$ everywhere, the Znajek regularity condition will require that $H_\varphi$ vanish at the event horizon.
I.e., either $C=0$ or $\Lambda$ diverges at the event horizon since $\Lambda \approx {\rm const}/\Delta$ near the event horizon. If $C=0$, we do not have a ``real" valued  $H_\varphi$. On the other hand if $\Lambda$ diverges at the event horizon, our solution becomes invalid at the event horizon (see expression for $E_r$ above). I.e., either way, our solution is not valid at the event horizon.

It turns out that the symmetry axis of the Kerr geometry is problematic for our new solution as well. For the Maxwell tensor to be well defined along the poles ($\theta = 0, \pi$) we must have that $\sqrt{\gamma} B^\theta /\sin\theta$ must be finite along the axis. This is not true in our case. Therefore, the solution presented here is only valid away from the poles and the event horizon of the black hole.
\begin{widetext}
Incidently, in the exterior geometry, the solution here is magnetically dominated since
$$B^2 - E^2 = g_{\theta \theta} (B^\theta)^2 + g^{\varphi \varphi} (B_\varphi)^2 - g^{rr} (E_r)^2 >  g_{\theta \theta} (B^\theta)^2  - g^{rr} (E_r)^2$$
$$=\frac{\Lambda^2 (\Omega_{,r})^2}{(r^2+a^2)^2\gamma}\left [\rho^2(r^2+a^2)^2 - \Sigma^2 a^2 \sin^2 \theta \right] >\frac{\Lambda^2 (\Omega_{,r})^2}{\gamma}(r^2-a^2)>0.$$
\end{widetext}
It would appear that the force-free solution we have obtained here by mere observation has limited utility. However, there are other interesting features contained in this solution. To see this, we first compute the current density vector. When $\Omega = \Omega_2$, the current density vector is given by
$$I = \frac{F^\prime}{\rho^2 \sin \theta} \left (\sin \theta,\; 0,\; \frac{F }{a^2 H_\varphi},\; \frac{1}{a \sin \theta}\right),$$
where $F= \Lambda \Delta \Omega \Omega_{,r}$ and $F^\prime = dF/dr$.
Then
$$I^2 = g(I,I) = \frac{(F^\prime)^2}{a^4\rho^2 \sin^2 \theta}  \left(1+ \frac{F^2}{C^2-F^2}\right)$$
which means that the current vector is spacelike since
$$C^2-F^2 = (a H_\varphi)^2.$$
As was mentioned above, we had previously obtained solutions where the current density vector was proportional to the principal geodesics $n,l$ of the Kerr geometry. Here, it turns out that the current vector is related to the remaining two vectors in the canonical null tetrad of the Kerr geometry, specifically, $m$ and $m^\star$, where
$$m = \frac{1}{\sqrt{2} \rho} \left(i a \sin \theta, \;0, \;1,\; \frac{i}{\sin \theta}\right)\;.$$
Here $\rho = r + i a \cos\theta$ (this is consistent with the notation that $\rho^2 = \rho \bar \rho \equiv (r + i a \cos\theta)(r -i a \cos\theta)$).
Clearly then, the current density vector above can be written in terms of the real and imaginary parts of $m$ as follows:
$$I = \frac{F^\prime}{a\rho^2 \sin \theta} \left[ {\rm Im} \;(\sqrt{2} \rho m) + \frac{F}{a H_\varphi} {\rm Re} \;(\sqrt{2} \rho m) \right].$$
The associated dual solution described in general in \cite{MD11} is generated by the same $\Omega_2$ and has a current vector of the form:
$$\tilde I = \frac{F^\prime}{a\rho^2 \sin \theta} \left[  {\rm Im} \;(\sqrt{2} \rho m) -\frac{F}{a H_\varphi} {\rm Re} \;(\sqrt{2} \rho m) \right].$$
In this case, the $\pm$ sign in eq.(\ref{rdephphi}) is exactly what distinguishes a solution from its dual.

It is indeed remarkable that there exists, four linearly independent solutions to the time-independent, stationary, axis-symmetric, force-free magnetosphere in a Kerr geometry where  the current density vectors are proportional to $n$ and $l$, and two other albeit local solutions proportional the linear combinations of the remaining bases vectors $m$ and $m^\star$ in the null tetrad of the Newman-Penrose formalism. Since $B^r = 0$ in the new solutions presented here (solutions generated by $\Omega_2$), they do not allow an exchange of energy and or angular momentum with the Black hole. This feature is an expected one from Christodoulou's general calculations since at the event horizon $\Omega_2 |_{r_+} = a/(r_+ ^2 + a^2) \equiv \Omega_H$,  the angular velocity of the event horizon.

\bibliography{bibliography}

\end{document}